\pdfoutput=1
\documentclass[runningheads]{llncs}
\usepackage{amsmath,amssymb,amsfonts}
\usepackage{graphicx}
\usepackage{caption}
\usepackage{multirow}
\usepackage{wrapfig}
\usepackage{hhline}
\usepackage{todonotes}
\usepackage{xcolor}
\definecolor{sig}{RGB}{159,255,180}

\definecolor{darkgreen}{RGB}{10,100,40}

\usepackage{color}
\usepackage[normalem]{ulem}

\usepackage{enumitem}
\setlist{label=\textbullet,leftmargin=*,wide=0pt}
\usepackage{tikz}
\usetikzlibrary{calc,arrows,positioning,shapes,shadows,spy,snakes,plotmarks,matrix,fit,backgrounds}

\begin{document}
\title{A Deep Network for Joint Registration and Reconstruction of Images with Pathologies}
\titlerunning{Network for Joint Registration and Reconstruction}
%
\author{Xu Han\inst{1}, Zhengyang Shen\inst{1}, Zhenlin Xu\inst{1}, Spyridon Bakas\inst{2}, Hamed Akbari\inst{2}, Michel Bilello\inst{2}, Christos Davatzikos\inst{2}, Marc Niethammer\inst{1}}
\authorrunning{Han et al.}
%
\institute{Department of Computer Science, UNC Chapel Hill, NC, USA \and Center for Biomedical Image Computing and Analytics, Perelman \\School of Medicine, University of Pennsylvania, PA, USA}
\maketitle              
\begin{abstract}
Registration of images with pathologies is challenging due to tissue appearance changes and missing correspondences caused by the pathologies. Moreover, mass effects as observed for brain tumors may displace tissue, creating larger deformations over time than what is observed in a healthy brain. Deep learning models have successfully been applied to image registration to offer dramatic speed up and to use surrogate information (e.g., segmentations) during training. However, existing approaches focus on learning registration models using images from healthy patients. They are therefore not designed for the registration of images with strong pathologies for example in the context of brain tumors, and traumatic brain injuries. In this work, we explore a deep learning approach to register images with brain tumors to an atlas. Our model learns an appearance mapping from images with tumors to the atlas, while simultaneously predicting the transformation to atlas space. Using separate decoders, the network disentangles the tumor mass effect from the reconstruction of quasi-normal images. Results on both synthetic and real brain tumor scans show that our approach outperforms cost function masking for registration to the atlas and that reconstructed quasi-normal images can be used for better longitudinal registrations.
\end{abstract}
\section{Introduction}
Registration is a fundamental problem in medical image analysis~\cite{sotiras2013deformable}. It aims at finding spatial correspondences between two images that are useful for many tasks, e.g., for atlas-based segmentation~\cite{aljabar2009multi}. Particularly for patients with brain tumors, an accurate image registration between the pre-operative and the post-recurrence images can help analyze the characteristics of tissue resulting in tumor recurrence~\cite{provenzale2006diffusion,kwon2013portr,akbari2016imaging,han2018patient}. Traditionally, image registration is formulated as an optimization problem seeking to minimize the dissimilarity between a warped source image and a target image while simultaneously encouraging spatially regular transformations. To capture large deformations, fluid-based registration models are frequently used~\cite{modersitzki2004numerical}, e.g., stationary velocity field (SVF)~\cite{vercauteren2009diffeomorphic} or large deformation diffeomorphic metric mapping (LDDMM) approaches~\cite{beg2005computing,avants2009advanced}, which can guarantee diffeomorphic transformations if sufficiently regularized. Non-parametric image registration models~\cite{modersitzki2004numerical} such as SVF and LDDMM require optimizing over millions of parameters in 3D, which is usually very slow. Hence, deep learning (DL) approaches have been proposed for such registration models~\cite{yang2017quicksilver,balakrishnan2019voxelmorph,shen2019networks,shen2019region}. By shifting the computational cost to the training time, DL approaches are orders of magnitudes faster at test time than numerical optimization, while retaining registration accuracy.

While many registration approaches for normal images or images with similar appearance have been proposed, a limited body of literature exists for the registration of images with pathologies, which is challenging due to tissue appearance differences and missing correspondences. Possible approaches include a) cost-function masking~\cite{brett2001spatial} (masking out tumor regions when calculating the similarity measure), b) use of robust similarity measures~\cite{reuter2010highly}, or c) replacing the pathology with quasi-normal appearance~\cite{liu2014low,yang2016registration,han2017efficient,han2018brain}. Masking out the tumor requires an accurate segmentation of the tumor region, and if it is large or in anatomically critical locations cost function masking may hide too much of the underlying brain structure, which should guide the registration~\cite{yang2016registration}. Reconstruction of quasi-normal appearance, on the other hand, does not require a prior segmentation and tumor-to-quasi-normal appearance can be learned via quasi-lesions with a variational autoencoder~\cite{yang2016registration}, or from a statistical model of a healthy population~\cite{liu2014low,han2017efficient,han2018brain}. The quasi-lesion approach~\cite{yang2016registration} introduces synthetic tumors and learns to reconstruct the underlying normal appearance, but the resulting reconstructions are still subject to mass effects and therefore do not properly disentangle appearance from such deformation changes. Existing approaches based on statistical models require underlying registrations to a common space for quasi-normal image reconstruction. But as a good alignment in cases of mass effect cannot be obtained without reconstruction, registration and reconstruction need to be interleaved in a costly iterative scheme.

A conceptually attractive approach would be to separate the mass effect and appearance changes and to reconstruct quasi-normal images in an atlas space, where appearance variability is expected to be lower. Inspired by a previous work on shape and appearance disentangling~\cite{shu2018deforming}, we propose a deep neural network to simultaneously register a brain tumor\footnote{Our goal is to register images with strong pathologies, e.g., tumors, traumatic brain injuries, or strokes. We focus on tumors in this work, but our approach is general.} image to an atlas while reconstructing a quasi-normal image {\it in atlas space}. The reconstructed quasi-normal image is in turn used in the similarity loss to guide our network to learn the spatial transformation from the image to the atlas.

\textbf{Contributions.} 1)\emph{Joint reconstruction and registration network.} To the best of our knowledge, this is the first deep network trained jointly to reconstruct and register brain images with strong pathologies to an atlas. The network recovers the missing correspondences between the pathologies and the atlas space. Our approach is also more computationally efficient than previous approaches by avoiding the interleaving of registrations and reconstructions, resulting in rapid predictions at test time. 2)\emph{Reconstruction of quasi-normal appearance in atlas space.} As we disentangle the transformation to the atlas from the reconstruction, we obtain tumor-to-quasi-normal image appearances in atlas space, thereby simplifying the appearance modeling. 3)\emph{Vector-momentum parameterized fluid-based registration.} Our network incorporates a vector-momentum parameterized stationary velocity field (vSVF)~\cite{shen2019networks}, which can capture large deformations while retaining diffeomorphic transformations. We use the reconstructed quasi-normal image to drive the registration, instead of the input tumor image. 4)\emph{Validation.} We show that our network successfully learns to reconstruct quasi-normal appearance simultaneously with the transformation of the tumor image to atlas space. Specifically, we show improvements over cost function masking, demonstrating that modeling quasi-normal image structure is beneficial for the registration of images with pathologies.

\textbf{Organization.} Sec.~\ref{sec:method} describes our registration and reconstruction network. Sec.~\ref{section:experiments} presents experimental details and results on both a synthetic brain tumor dataset and on paired sets of pre-operative and post-recurrence brain tumor scans. Sec.~\ref{section:conclusion} concludes with a summary and an outlook on future work.

\section{Methodologies}
\label{sec:method}
\begin{figure}[b]
	\includegraphics[width=\textwidth]{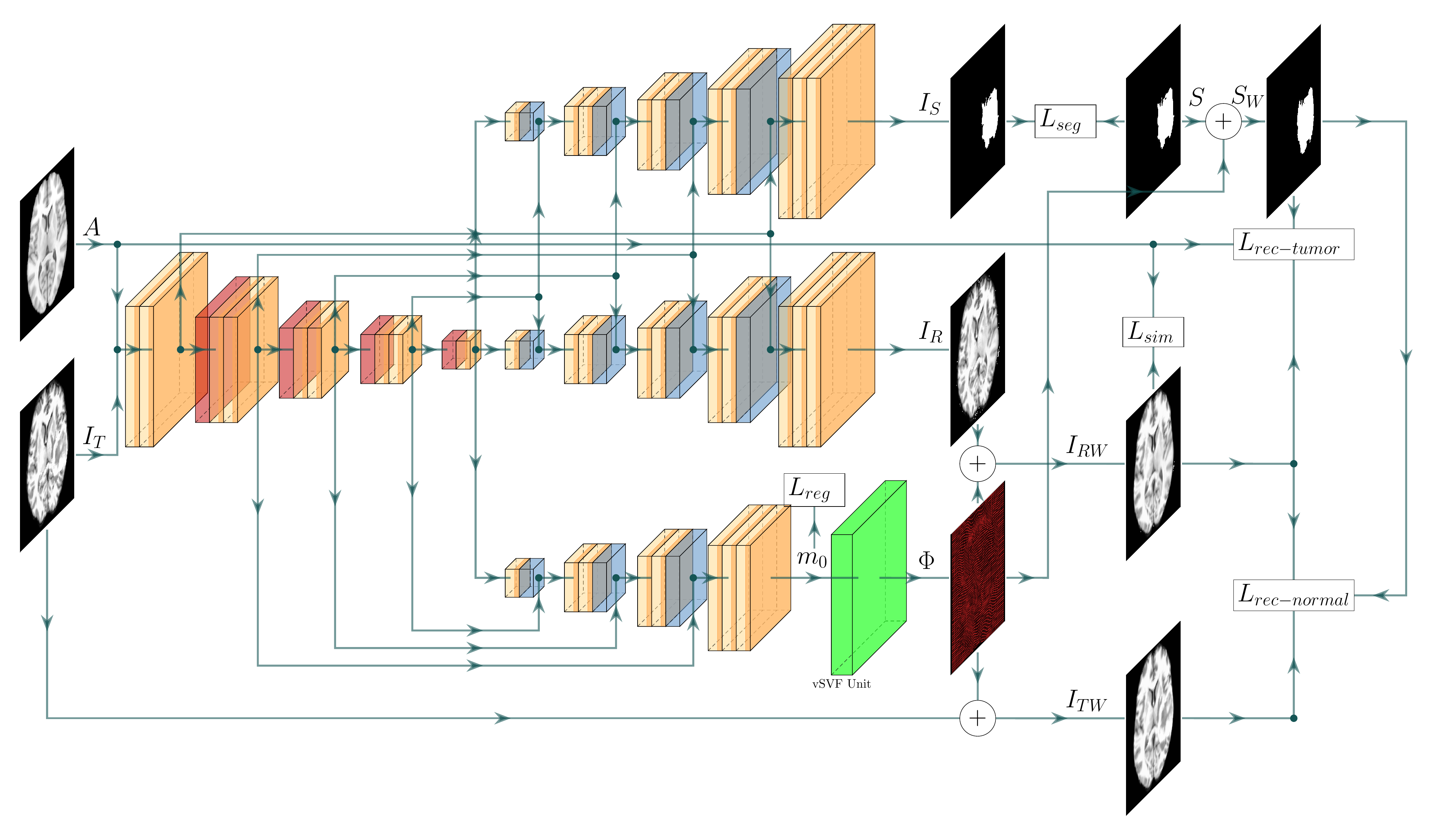}
	\caption{Overview of the proposed network. The network outputs a mask $I_S$, a reconstructed quasi-normal image $I_R$ and a vector momentum $m_0$ which is used to obtain the transformation map $\Phi^{-1}$. The regularization loss $L_{reg}$ penalizes $m_0$, while the similarity loss $L_{sim}$ penalizes the warped reconstructed image $I_{RW}$ with respect to the atlas $A$. The reconstruction loss penalizes the warped quasi-normal image in the tumor region and the normal region.}
	\label{fig:framwork}
\end{figure}

This section describes our deep network, including its architecture and the associated loss functions. Fig.~\ref{fig:framwork} shows an overview of our network. The network takes a tumor image $I_T$ and an atlas $A$ as its inputs and outputs a vector-momentum parameterization of the transformation $\Phi^{-1}$, a reconstructed quasi-normal image $I_R$ and a segmentation of the tumor region $I_S$. The network jointly learns both the registration and reconstruction, which is more efficient than approaches that interleave registrations and reconstructions. Importantly, the transformation warps the tumor image to the atlas for a better reconstruction in {\em atlas space}, while the {\it reconstructed image} guides the similarity measure so that the network learns a better transformation as it is no longer perturbed by the pathology.

\textbf{Registration.} We use a vector-momentum
parameterized stationary velocity field (vSVF) model~\cite{shen2019networks,niethammer2019metric}. Instead of directly predicting the transformation field, our network predicts a momentum vector field, $m$, which gets smoothed by a multi-Gaussian kernel~\cite{risser2011simultaneous} resulting in a velocity vector field, $v$, from which the transformation map, $\Phi^{-1}$, is computed via integration. The benefit of this indirect way is that it can assure diffeomorphic transformations at {\em test} time.

The registration loss consists of a regularization loss and a similarity loss:
\begin{equation}
\label{eq:svf}
\begin{split}
L_{rgs}(m_0) =& \langle m_0, v_0 \rangle + \frac{1}{\sigma^2} Sim[I_R \circ \Phi^{-1}(1), A],\quad \\
&\Phi_t^{-1} + D\Phi^{-1}v_0 = 0,~\text{s.t.}\quad\Phi^{-1}(0) = \Phi_{(0)}^{-1},\quad v_0 = (L^\dagger L)^{-1}m_0,
\end{split}
\end{equation}
where $D$ denotes the Jacobian, $m_0$ is the initial vector momentum, $\sigma>0$ balances the two terms, and $\Phi^{-1}_{(0)}$ is the initial condition for the transformation map, $\Phi^{-1}$, which can be set to identity or to the transformation of a pre-registration, for example, an affine registration. $\|v\|_L^2 = \langle L^{\dagger}Lv, v\rangle$ is a norm defined by a differential operator $L$ and its adjoint $L^{\dagger}$~\cite{beg2005computing}. We use localized normalized cross correlation (LNCC) as our similarity loss as in~\cite{shen2019networks}. A significant difference from existing registration networks is that instead of using the input tumor image $I_T$ to evaluate the similarity loss, we use the reconstructed image $I_R$. The reconstructed $I_R$ recovers image correspondences which can guide image registration. The registration loss only backpropagates through the registration decoder.

\textbf{Reconstruction.}
The reconstruction decoder predicts a quasi-normal image from the tumor image. We directly learn this mapping from the atlas appearance. Specifically, for a given tumor image, we use its manually segmented tumor mask, $S$, to separate the tumor and the normal region. The tumor mask is only used during training. In the normal region, the warped reconstruction image $I_{RW} = I_R \circ \Phi^{-1}(1)$ should be close to the warped original image $I_{TW} = I_T \circ \Phi^{-1}(1)$. In the tumor region, the reconstruction should be close to the atlas $A$. The warped tumor mask is $S_W = S \circ \Phi^{-1}(1)$. We define the reconstruction loss as follows: 

\begin{equation}
L_{rec} = \frac{1}{|\Omega_N|}\int_{\Omega_N} (I_{RW} - I_{TW})^2~dx + \frac{1}{|\Omega_T|}\int_{\Omega_T} (I_{RW} - A)^2~dx,
\end{equation}
where $\Omega_N = \{x:S_W(x)=0\}$ is the normal domain, $\Omega_T = \{x:S_W(x)=1\}$ is the tumor domain, and $|\Omega|$ denotes the volume of domain $\Omega$. The loss captures the sum of the mean-squared errors over the normal region and the tumor region. We use atlas appearance to learn the tumor-to-quasi-normal mapping since the atlas is our target image. This can be considered a highly simplified statistical model only represented by its mean, the atlas. Combinations with more advanced statistical models, for example based on principal component analysis~\cite{liu2014low,han2017efficient} or variational autoencoders~\cite{kingma2013auto}, are conceivable. The reconstruction loss only backpropagates through the reconstruction decoder.

\textbf{Segmentation.} In principle, the segmentation decoder is not required for registration and reconstruction. Since we use the segmentation mask during training for reconstruction, we also add a segmentation decoder which outputs a predicted segmentation of the tumor. This is similar to~\cite{shu2018deforming}, where an instance class can also be predicted. Intuitively, by providing direct supervision on the segmentation, the network is required to learn a representation capable of separating the tumor from the normal region. We use binary cross-entropy loss, where the output of the segmentation decoder is the predicted probability that a voxel belongs to the tumor region: $L_{seg} = Bce[I_S, S]$.

\section{Experiments and Results}
\label{section:experiments}

We created a pseudo-tumor dataset providing us with a synthetic ground-truth for the reconstructions. We show that it is beneficial to use our quasi-normal image reconstructions for registration. We also use a dataset of pre-operative and post-recurrence magnetic resonance images (MRIs) from patients with glioblastomas with expert-placed landmarks for validation. We show that the predicted registration by our network is more accurate than cost function masking and direct registration of the tumor images. We use ICBM 152~\cite{fonov2009unbiased} as our atlas.

\noindent
\textbf{3D Pseudo-tumor.} 
We created this dataset using BraTS2019~\cite{menze2014multimodal,bakas2017advancing,bakas2018identifying} and OASIS-3~\cite{lamontagne2018oasis}. OASIS-3 contains longitudinal MRIs from over 1,000 participants with normal cognitive function and with various stages of cognitive decline. The BraTS data contains MRIs from patients with brain tumors and corresponding tumor segmentations. We randomly selected 280 pairs of T1w-images; one from OASIS (we only use one scan for each patient) and one from BraTS. To mimic the mass effect of the brain tumor, we registered the OASIS T1w scan to the BraTS T1w scan with cost function masking and pasted the brain tumor from the BraTS scan onto the deformed OASIS scan. The resulting 280 simulated images are our pseudo-tumor dataset. We randomly select 40 for testing, 40 for validation and 200 for training. Images are affinely aligned to the atlas, which is resampled to $128\times128\times128$ with $1.5\times1.5\times1.5$~mm$^3$ isotropic voxels.  

Since this dataset is simulated, we have the images without the added tumor but including the spatial transformation. We register the atlas to these images. As these registrations are not impacted by the tumors, but might not reflect the exact correspondence (due to possible registration errors), we regard the resulting registrations as the gold-standard to which we compare in the following. We register the atlas to: 1) the tumor images (\texttt{TUMOR}), 2) the tumor images using cost function masking, 3) the quasi-normal images predicted by a network with a quasi-lesion layer~\cite{yang2016registration}(\texttt{REC\_QL}) and 4) the quasi-normal images predicted by our network. As the gold-standard is obtained through optimization, we perform all the registrations using the same \emph{optimization} model for the pseudo-tumor dataset and do not compare to the predicted registrations. For cost function masking, we conduct two experiments using different masks, one using the groundtruth masks (\texttt{CFM\_GM}) and one using the predicted masks by our network (\texttt{CFM\_PM}). Using the predicted masks (\texttt{CFM\_PM}) allows us to evaluate the performance of cost function masking, when groundtruth (or manually segmented) masks are not available at test time, which is often the case. For our model, we train with (\texttt{REC\_RRS}) and without (\texttt{REC\_RR}) the segmentation decoder. In addition, for the predicted quasi-normal images, we can keep the normal tissue unchanged by using the predicted segmentation (\texttt{REC\_RRS\_PM}). We compare the deformation differences between the results obtained by each of the optimization-based registrations and our gold standard registration result.

\begin{table}[b]
    \centering
    \begin{tabular}{@{}c@{}c@{}c@{}}
        \includegraphics[width=0.30\columnwidth]{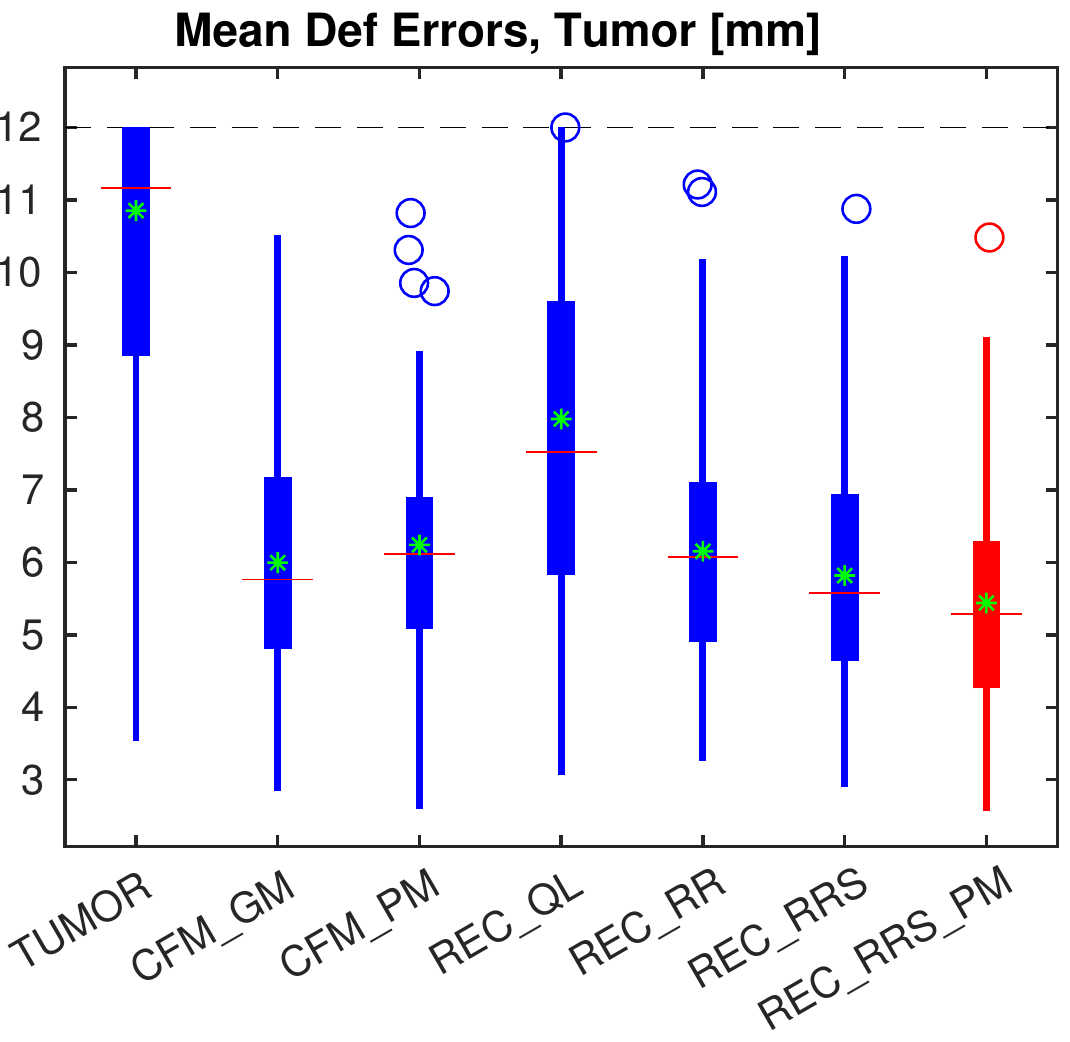} &
        \includegraphics[width=0.30\columnwidth]{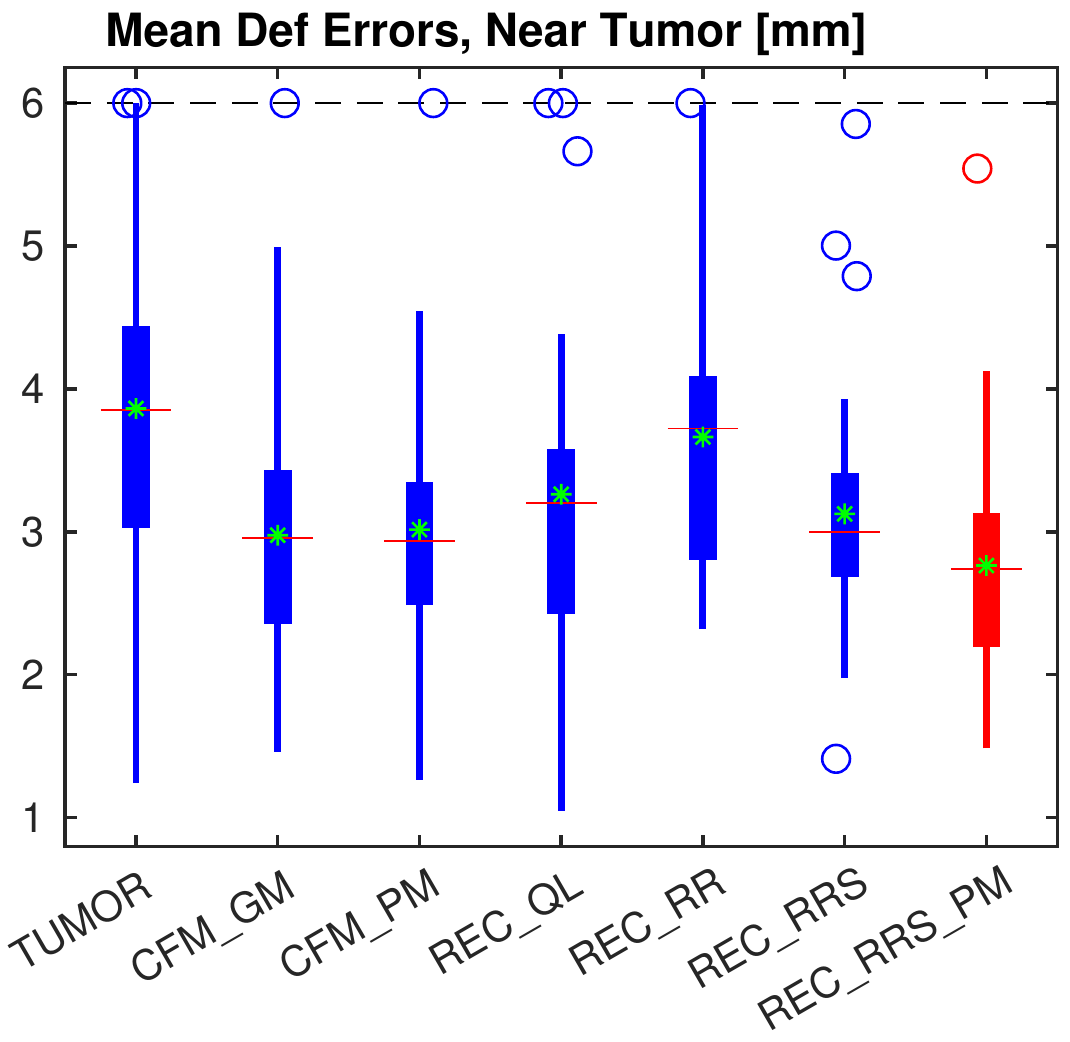} &
        \includegraphics[width=0.30\columnwidth]{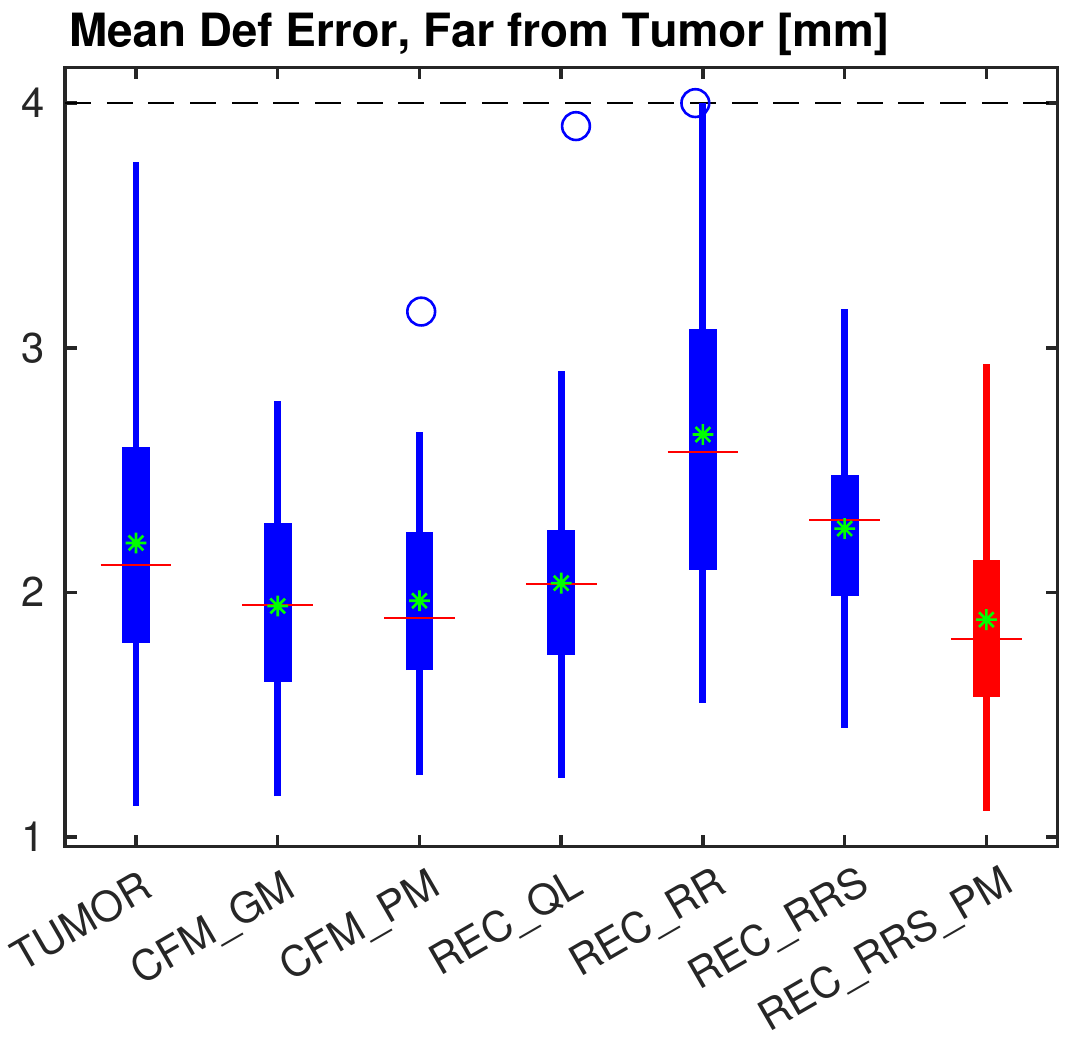}
    \end{tabular}
    \captionof{figure}{Boxplots of mean deformation differences with respect to the gold standard deformations. 
    \texttt{TUMOR}: directly registering to the tumor image; \texttt{CFM}: cost function masking, where \texttt{\_GM} and \texttt{\_PM} refer to using the groundtruth masks and predicted masks, respectively; \texttt{REC\_*}s: registering to the reconstructed images, where \texttt{REC\_QL} uses the quasi-lesion layer, \texttt{REC\_RR} only uses the registration and reconstruction decoders. \texttt{REC\_RRS}: proposed network using registration, reconstruction, and segmentation decoders. In addition, \texttt{REC\_RRS\_PM} (in red) retains the normal region in areas predicted by the masks obtained by our network.}
    \label{fig:box_pseudo}
\end{table}

Fig.~\ref{fig:box_pseudo} shows the results for the pseudo-tumor dataset. For each case, we evaluate the mean deformation differences in three regions: 1) the tumor region, 2) the normal region near the tumor (within 30~mm), and 3) the normal region far from the tumor (over 30~mm). Our network performs much better when the segmentation decoder is used, because of the additional supervision (\texttt{REC\_RRS vs. REC\_RR}). The network using the quasi-lesion layer (\texttt{REC\_QL}) works well in the normal region but performs poorly in the tumor region. This might be because at test time the real tumor region is subject to larger mass effects than what was captured during training, as quasi-lesions can never be introduced inside the actual tumor region. Compared to cost function masking, our method (\texttt{REC\_RRS\_PM}), on average, improves by about 0.5~mm in the tumor region when the groundtruth masks are available (\texttt{CFM\_GM}) and around 0.8~mm when the groundtruth masks are not available (\texttt{CFM\_PM}). In the normal regions, improvements over cost function masking are relatively small, around 0.3~mm.

\noindent
\textbf{3D Real Brain Tumor.}
This dataset consists of images for 22 patients with brain glioblastoma. Each patient has scans from two time-points, one before the surgery (pre-operative) and one after surgery (post-recurrence). All images are of size $155\times240\times240$ with isotropic voxels $1\times1\times1$ mm$^3$. We only use the T1w images in the dataset. For each patient, a radiologist placed 10 landmarks near the tumor (within 30~mm) and 10 landmarks far from the tumor (over 30~mm) in both the pre- and post-scans. We train our network using a subset of the BraTS2019 training data with 120 training images and 20 validation images. Testing is performed via our glioblastoma dataset. To limit dataset variability, we selected a subset of the BraTS training data, which was acquired by one institution and which is similar in acquisition to our test data. Ideally, our test dataset is used for longitudinal registration, i.e., registering between the pre- and post-scans from the same patient. As our network predictions are with respect to an atlas we conduct the following two experiments:

\begin{itemize}
\item \textbf{Atlas Registration.}
For each patient, we feed both scans into our network and obtain respective transformations to the atlas. We then compose the forward map of the pre-scan and the inverse map of the post-scan, resulting in a pre-atlas-post (\texttt{REC\_PAP}) map. To compare, we also perform an optimization-based atlas-registration directly using the tumor images (\texttt{TUMOR\_PAP}) and with cost function masking (\texttt{CFM\_PAP}). In both cases, we obtain the composited transformation. Using the resulting transformations, we warp the landmarks from the post-scan to the pre-scan space and evaluate the landmark differences. As we do not have manual tumor segmentations, we use predicted masks for cost function masking.

\begin{table}[b]
    \centering
    \begin{tabular}{@{}c@{}c@{}}
        \includegraphics[width=0.5\columnwidth]{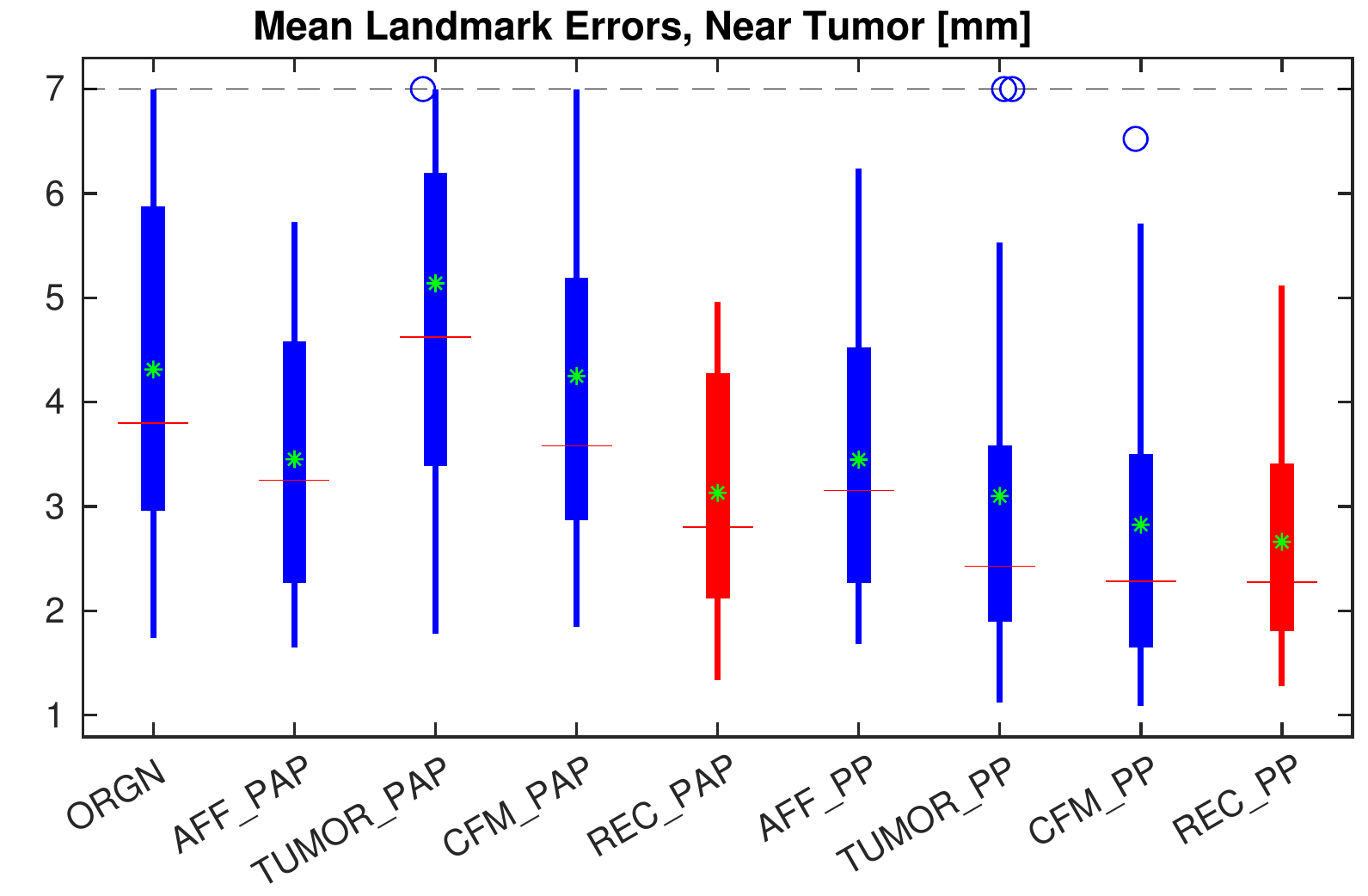} &
        \includegraphics[width=0.5\columnwidth]{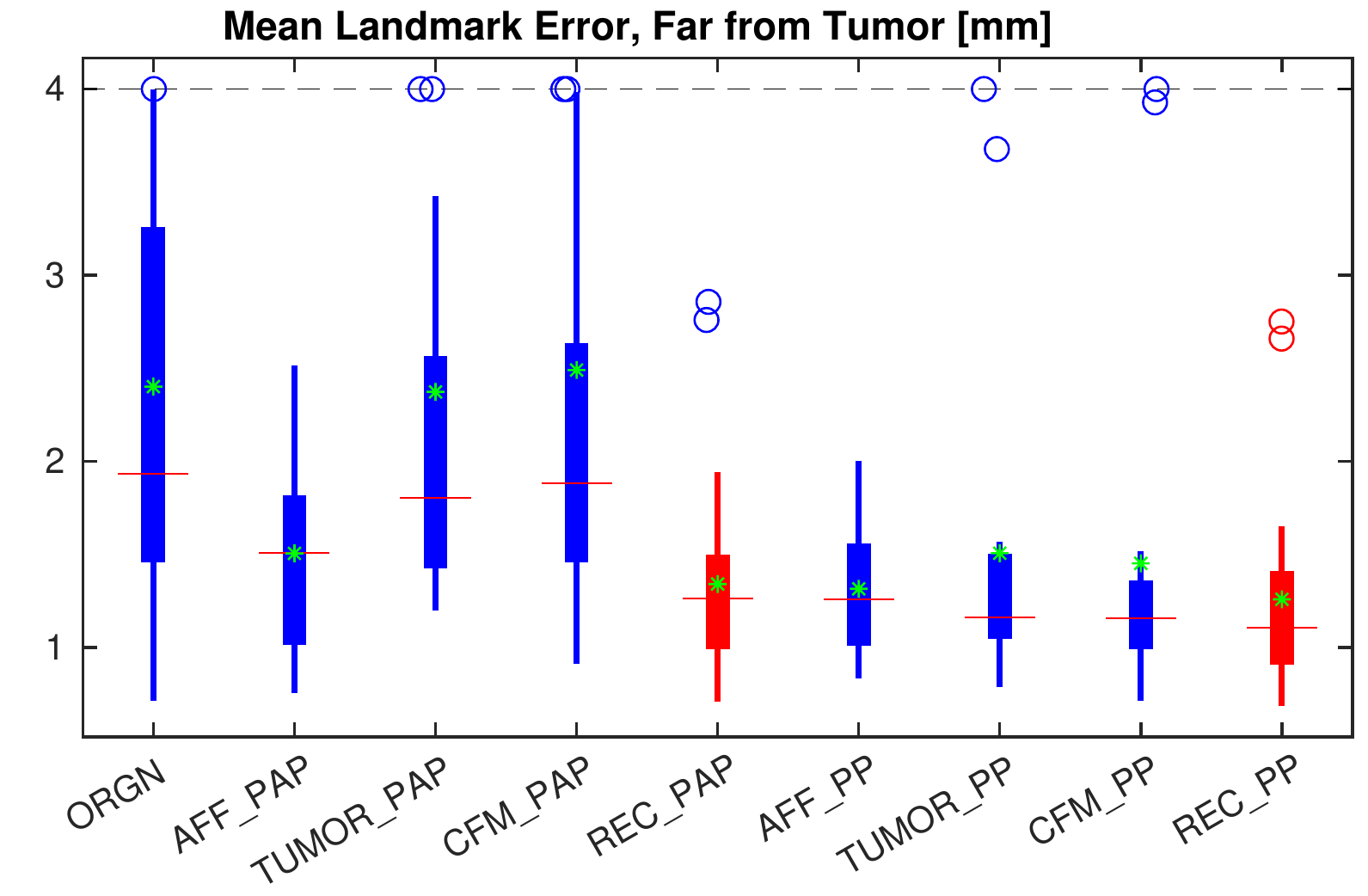}
    \end{tabular}
    \captionof{figure}{Boxplots of mean landmark errors for registration of glioblastoma patients. (\texttt{ORGN}) is the landmark differences before registration. The next four are results via the atlas, i.e, pre-atlas-post (\texttt{\_PAP}); the last four are longitudinal results, i.e., pre-post (\texttt{\_PP}). We compare to affine registration, registration of tumor images, and cost function masking.}
    \label{fig:box_real}
\end{table}

\item \textbf{Longitudinal Registration.} 
We perform optimization-based vSVF registrations between reconstructed quasi-normal images of both the scans, predicted by our network (\texttt{REC\_PP}). We compare with longitudinally registering directly using tumor images (\texttt{TUMOR\_PP}) and using cost function masking (\texttt{CFM\_PP}).
\end{itemize}

Fig.~\ref{fig:box_real} shows landmark errors in two different regions for the different registration approaches. When registrations are composed through the atlas, errors are much larger than direct longitudinal registration. However, our method shows improvements over cost function masking in both cases. Finally, Fig.~\ref{fig:tumor_example} shows an example for a brain tumor image. The 3rd column is the predicted quasi-normal image, and the 4th column is the warped image in atlas space. We observe some contrast differences between the tumor and the normal region. However, as our goal is registration, it is not an issue as the correspondences can be established between the reconstructed image and the atlas.

\begin{table}[b]
    \setlength\tabcolsep{-0.01\columnwidth}
    \centering
    \begin{tabular}{ccccc}
        \begin{tikzpicture}[thick, spy using outlines={rectangle,lens={scale=2}, size=1.5cm, connect spies}]
	    \node{\includegraphics[width=0.2\columnwidth]{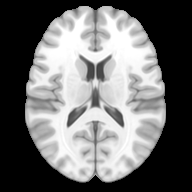}};%
            \spy [green] on (0,0.4) in node [left] at (1.2,2.0);%
          \end{tikzpicture} &
        \begin{tikzpicture}[thick, spy using outlines={rectangle,lens={scale=2}, size=1.5cm, connect spies}]
	    \node{\includegraphics[width=0.2\columnwidth]{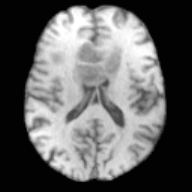}};%
            \spy [green] on (0,0.4) in node [left] at (1.2,2.0);%
          \end{tikzpicture} &
        \begin{tikzpicture}[thick, spy using outlines={rectangle,lens={scale=2}, size=1.5cm, connect spies}]
	    \node{\includegraphics[width=0.2\columnwidth]{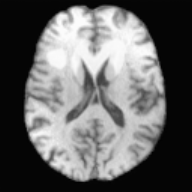}};%
            \spy [green] on (0,0.4) in node [left] at (1.2,2.0);%
          \end{tikzpicture} & 
        \begin{tikzpicture}[thick, spy using outlines={rectangle,lens={scale=2}, size=1.5cm, connect spies}]
	    \node{\includegraphics[width=0.2\columnwidth]{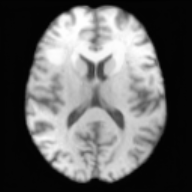}};%
            \spy [green] on (0,0.4) in node [left] at (1.2,2.0);%
          \end{tikzpicture} &
        \begin{tikzpicture}[thick, spy using outlines={rectangle,lens={scale=2}, size=1.5cm, connect spies}]
	    \node{\includegraphics[width=0.2\columnwidth]{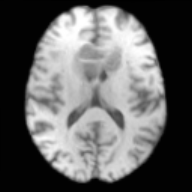}};%
            \spy [green] on (0,0.4) in node [left] at (1.2,2.0);%
          \end{tikzpicture}\\
        $A$&$I_T$&$I_R$&$I_{RW}$&$I_{TW}$
    \end{tabular}
    \captionof{figure}{One example network result for a brain tumor image. The 5 columns show: 1) the atlas; 2) the tumor image; 3) the reconstructed quasi-normal image, predicted by our network; 4) the warped quasi-normal image; and 5) the warped tumor by applying the transformation.}
    \label{fig:tumor_example}
\end{table}

\section{Conclusion}
\label{section:conclusion}
	In this work, we proposed a joint registration and reconstruction network. Given a brain image with pathologies, our network simultaneously learns a registration to a common atlas space and a reconstruction of quasi-normal appearance in the atlas space. Our experiments show that, as the network disentangles the spatial variation (e.g., caused by mass effects) from the appearance differences of the pathology, the reconstructed quasi-normal appearance provides better guidance to the registration. Future work could incorporate statistical models based on principal component analysis to capture appearance variations in atlas space. 

\section*{Acknowledgments}
Research reported in this publication was supported by the National Institutes of Health (NIH) and the National Science Foundation (NSF) under award numbers NINDS:R01NS042645, NCI:U24CA189523/U01CA242871, NSF:EECS1711776, and NIH:1R21CA22330401. The content is solely the responsibility of the authors and does not represent the official views of the NIH or the NSF.

%


\bibliographystyle{splncs04}
\bibliography{ref1}

\begin{thebibliography}{10}
\providecommand{\url}[1]{\texttt{#1}}
\providecommand{\urlprefix}{URL }
\providecommand{\doi}[1]{https://doi.org/#1}

\bibitem{akbari2016imaging}
Akbari, H., Macyszyn, L., Da, X., Bilello, M., Wolf, R.L., Martinez-Lage, M.,
  Biros, G., Alonso-Basanta, M., O'Rourke, D.M., Davatzikos, C.: Imaging
  surrogates of infiltration obtained via multiparametric imaging pattern
  analysis predict subsequent location of recurrence of glioblastoma.
  Neurosurgery  \textbf{78}(4),  572--580 (2016)

\bibitem{aljabar2009multi}
Aljabar, P., Heckemann, R.A., Hammers, A., Hajnal, J.V., Rueckert, D.:
  Multi-atlas based segmentation of brain images: atlas selection and its
  effect on accuracy. Neuroimage  \textbf{46}(3),  726--738 (2009)

\bibitem{avants2009advanced}
Avants, B.B., Tustison, N., Song, G.: Advanced normalization tools ({ANTS}).
  Insight j  \textbf{2}(365),  1--35 (2009)

\bibitem{bakas2017advancing}
Bakas, S., Akbari, H., Sotiras, A., Bilello, M., Rozycki, M., Kirby, J.S.,
  Freymann, J.B., Farahani, K., Davatzikos, C.: Advancing the cancer genome
  atlas glioma {MRI} collections with expert segmentation labels and radiomic
  features. Scientific data  \textbf{4},  170117 (2017)

\bibitem{bakas2018identifying}
Bakas, S., Reyes, M., Jakab, A., Bauer, S., Rempfler, M., Crimi, A., Shinohara,
  R.T., Berger, C., Ha, S.M., Rozycki, M., et~al.: Identifying the best machine
  learning algorithms for brain tumor segmentation, progression assessment, and
  overall survival prediction in the {BRATS} challenge. arXiv preprint
  arXiv:1811.02629  (2018)

\bibitem{balakrishnan2019voxelmorph}
Balakrishnan, G., Zhao, A., Sabuncu, M.R., Guttag, J., Dalca, A.V.:
  {VoxelMorph}: a learning framework for deformable medical image registration.
  IEEE transactions on medical imaging  \textbf{38}(8),  1788--1800 (2019)

\bibitem{beg2005computing}
Beg, M.F., Miller, M.I., Trouv{\'e}, A., Younes, L.: Computing large
  deformation metric mappings via geodesic flows of diffeomorphisms.
  International journal of computer vision  \textbf{61}(2),  139--157 (2005)

\bibitem{brett2001spatial}
Brett, M., Leff, A.P., Rorden, C., Ashburner, J.: Spatial normalization of
  brain images with focal lesions using cost function masking. Neuroimage
  \textbf{14}(2),  486--500 (2001)

\bibitem{fonov2009unbiased}
Fonov, V.S., Evans, A.C., McKinstry, R.C., Almli, C., Collins, D.: Unbiased
  nonlinear average age-appropriate brain templates from birth to adulthood.
  NeuroImage (47), ~S102 (2009)

\bibitem{han2018patient}
Han, X., Bakas, S., Kwitt, R., Aylward, S., Akbari, H., Bilello, M.,
  Davatzikos, C., Niethammer, M.: Patient-specific registration of
  pre-operative and post-recurrence brain tumor {MRI} scans. In: International
  MICCAI Brainlesion Workshop. pp. 105--114. Springer (2018)

\bibitem{han2018brain}
Han, X., Kwitt, R., Aylward, S., Bakas, S., Menze, B., Asturias, A., Vespa, P.,
  Van~Horn, J., Niethammer, M.: Brain extraction from normal and pathological
  images: a joint {PCA}/image-reconstruction approach. NeuroImage
  \textbf{176},  431--445 (2018)

\bibitem{han2017efficient}
Han, X., Yang, X., Aylward, S., Kwitt, R., Niethammer, M.: Efficient
  registration of pathological images: a joint {PCA}/image-reconstruction
  approach. In: 2017 IEEE 14th International Symposium on Biomedical Imaging
  (ISBI 2017). pp. 10--14. IEEE (2017)

\bibitem{kingma2013auto}
Kingma, D.P., Welling, M.: Auto-encoding variational bayes. arXiv preprint
  arXiv:1312.6114  (2013)

\bibitem{kwon2013portr}
Kwon, D., Niethammer, M., Akbari, H., Bilello, M., Davatzikos, C., Pohl, K.M.:
  {PORTR}: Pre-operative and post-recurrence brain tumor registration. IEEE
  transactions on medical imaging  \textbf{33}(3),  651--667 (2013)

\bibitem{lamontagne2018oasis}
LaMontagne, P.J., Keefe, S., Lauren, W., Xiong, C., Grant, E.A., Moulder, K.L.,
  Morris, J.C., Benzinger, T.L., Marcus, D.S.: {OASIS-3}: Longitudinal
  neuroimaging, clinical, and cognitive dataset for normal aging and
  alzheimer’s disease. Alzheimer's \& Dementia: The Journal of the
  Alzheimer's Association  \textbf{14}(7),  P1097 (2018)

\bibitem{liu2014low}
Liu, X., Niethammer, M., Kwitt, R., McCormick, M., Aylward, S.: Low-rank to the
  rescue---atlas-based analyses in the presence of pathologies. In:
  International Conference on Medical Image Computing and Computer-Assisted
  Intervention. pp. 97--104. Springer (2014)

\bibitem{menze2014multimodal}
Menze, B.H., Jakab, A., Bauer, S., Kalpathy-Cramer, J., Farahani, K., Kirby,
  J., Burren, Y., Porz, N., Slotboom, J., Wiest, R., et~al.: The multimodal
  brain tumor image segmentation benchmark ({BRATS}). IEEE transactions on
  medical imaging  \textbf{34}(10),  1993--2024 (2014)

\bibitem{modersitzki2004numerical}
Modersitzki, J.: Numerical methods for image registration. Oxford University
  Press on Demand (2004)

\bibitem{niethammer2019metric}
Niethammer, M., Kwitt, R., Vialard, F.X.: Metric learning for image
  registration. In: Proceedings of the IEEE Conference on Computer Vision and
  Pattern Recognition. pp. 8463--8472 (2019)

\bibitem{provenzale2006diffusion}
Provenzale, J.M., Mukundan, S., Barboriak, D.P.: Diffusion-weighted and
  perfusion {MR} imaging for brain tumor characterization and assessment of
  treatment response. Radiology  \textbf{239}(3),  632--649 (2006)

\bibitem{reuter2010highly}
Reuter, M., Rosas, H.D., Fischl, B.: Highly accurate inverse consistent
  registration: a robust approach. Neuroimage  \textbf{53}(4),  1181--1196
  (2010)

\bibitem{risser2011simultaneous}
Risser, L., Vialard, F.X., Wolz, R., Murgasova, M., Holm, D.D., Rueckert, D.:
  Simultaneous multi-scale registration using large deformation diffeomorphic
  metric mapping. IEEE transactions on medical imaging  \textbf{30}(10),
  1746--1759 (2011)

\bibitem{shen2019networks}
Shen, Z., Han, X., Xu, Z., Niethammer, M.: Networks for joint affine and
  non-parametric image registration. In: Proceedings of the IEEE Conference on
  Computer Vision and Pattern Recognition. pp. 4224--4233 (2019)

\bibitem{shen2019region}
Shen, Z., Vialard, F.X., Niethammer, M.: Region-specific diffeomorphic metric
  mapping. In: Advances in Neural Information Processing Systems. pp.
  1096--1106 (2019)

\bibitem{shu2018deforming}
Shu, Z., Sahasrabudhe, M., Alp~Guler, R., Samaras, D., Paragios, N., Kokkinos,
  I.: Deforming autoencoders: Unsupervised disentangling of shape and
  appearance. In: Proceedings of the European Conference on Computer Vision
  (ECCV). pp. 650--665 (2018)

\bibitem{sotiras2013deformable}
Sotiras, A., Davatzikos, C., Paragios, N.: Deformable medical image
  registration: A survey. IEEE transactions on medical imaging  \textbf{32}(7),
   1153--1190 (2013)

\bibitem{vercauteren2009diffeomorphic}
Vercauteren, T., Pennec, X., Perchant, A., Ayache, N.: Diffeomorphic demons:
  Efficient non-parametric image registration. NeuroImage  \textbf{45}(1),
  S61--S72 (2009)

\bibitem{yang2016registration}
Yang, X., Han, X., Park, E., Aylward, S., Kwitt, R., Niethammer, M.:
  Registration of pathological images. In: International Workshop on Simulation
  and Synthesis in Medical Imaging. pp. 97--107. Springer (2016)

\bibitem{yang2017quicksilver}
Yang, X., Kwitt, R., Styner, M., Niethammer, M.: Quicksilver: Fast predictive
  image registration---a deep learning approach. NeuroImage  \textbf{158},
  378--396 (2017)

\end{thebibliography}


\end{document}